\documentclass[]{interact}

\usepackage{epstopdf}
\usepackage[caption=false]{subfig}

\usepackage[longnamesfirst,sort]{natbib}
\bibpunct[, ]{(}{)}{;}{a}{,}{,}
\renewcommand\bibfont{\fontsize{10}{12}\selectfont}

\theoremstyle{plain}

\theoremstyle{definition}

\theoremstyle{remark}

\begin{document}


\title{A Bayesian Mixture Model approach to expected possession values in rugby league}

\author{
\name{Thomas Sawczuk\textsuperscript{1,2}\thanks{CONTACT Thomas Sawczuk. Email: tsawczuk@hotmail.com}, Anna Palczewska\textsuperscript{1}, Ben Jones\textsuperscript{2,3} and Jan Palczewski\textsuperscript{4}}
\affil{\textsuperscript{1}School of Built Environment, Engineering and Computing, Leeds Beckett University, Leeds, United Kingdom; \textsuperscript{2}Carnegie Applied Rugby Research (CARR) Centre, Carnegie School of Sport, Leeds Beckett University, Leeds, United Kingdom; \textsuperscript{3}England Performance Unit, The Rugby Football League, Red Hall, Leeds, United Kingdom; \textsuperscript{4}School of Mathematics, Leeds University, Leeds, United Kingdom }
}

\maketitle

\begin{abstract}
The aim of this study was to improve previous zonal approaches to expected possession value (EPV) models in low data availability sports by introducing a Bayesian Mixture Model approach to an EPV model in rugby league. 99,966 observations from the 2021 Super League season were used. A set of 33 centres (30 in the field of play, 3 in the try area) were located across the pitch. Each centre held the probability of five possession outcomes occurring (converted/unconverted try, penalty, drop goal and no points). Weights for the model were provided for each location on the pitch using linear and bilinear interpolation techniques. Probabilities at each centre were estimated using a Bayesian approach and extrapolated to all locations on the pitch. An EPV measure was derived from the possession outcome probabilities and their points value. The model produced a smooth pitch surface, which was able to provide different possession outcome probabilities and EPVs for every location on the pitch. Differences between team attacking and defensive plots were visualised and an actual vs expected player rating system was developed. The model provides significantly more flexibility than previous approaches and could be adapted to other sports where data is similarly sparse.
\end{abstract}


\section{Introduction}
Across sport, the use of advanced statistical and machine learning methods to evaluate player and team performances through expected possession value (EPV) models is growing (\cite{Cervone2016, Fernandez2021, Liu2020, Decroos2019, Kempton2016, Sawczuk2021}). EPV models value every action and/or location on the field of play with respect to its point scoring potential. These models have become particularly prevalent in football (\cite{Fernandez2021, Liu2020, Decroos2019}) and basketball (\cite{Cervone2016}), where the quality (i.e. match event data and player tracking data) and quantity (i.e. millions of observations) of data allows practitioners to use them. In other sports, where significantly less data is available, it is not possible to adapt these advanced EPV models so different approaches must be considered (\cite{Kempton2016, Sawczuk2021}).

Rugby league is an example of a sport with low data availability. Despite its low data availability relative to other sports, two versions of an EPV model have been published within rugby league (\cite{Kempton2016, Sawczuk2021}). Both studies utilised a Markov Reward Process (MRP) approach (\cite{White1989}), which required data to be aggregated and valued equally within a set of zones. Aggregating data into zones is a common practice within spatial analyses across multiple domains (\cite{Lee2014, Stillwell2018}) and has been used successfully in basketball (\cite{Cervone2016}). Unfortunately, rugby league has a much larger playing surface than basketball, which limits the usefulness of this method when attempting to evaluate team and player performances. Indeed, the smallest zone size employed within either rugby league analysis to date had an area of 100m\textsuperscript{2}, which would cover almost a quarter of a professional basketball court. The impracticality of this situation was shown when Sawczuk et al. (\cite{Sawczuk2021-UKCI}) attempted to extend their EPV-19 model to provide player ratings and found that in a worst-case scenario, a player could run 59m forward on the pitch from their team’s 10m line to just short of the opposition 30m line and receive no positive value for their action. Conversely, moving 1m more central in the opposition 20m, whilst moving 19m backwards, could result in a large positive value being given to the player. A second limitation of the MRP approach is that it aggregates all scoring outcomes into a single value. In rugby league where there are five scoring outcomes, it is possible that this could result in the loss of valuable tactical information. For example, a team may be more likely to score converted tries on one side of the pitch, but unconverted tries on the other. Similarly, they may be more likely to score drop goals than penalty goals from a specific area on the pitch. If the probability of each scoring outcome could be modelled individually, before being combined to produce an EPV measure, it may be possible to glean specific tactical insights from the data. There is therefore scope for the adoption of a novel approach to EPV modelling, which allows a smooth pitch surface to be calculated for each possession outcome probability in a low data availability sport. 

One method through which the limitations of previous EPV models in rugby league could be solved is through a custom built Bayesian Mixture Model (\cite{Ullah2019}). The usefulness of a Bayesian approach to analysing spatial data has previously been shown in sport (\cite{Hobbs2020, Cervone2016}). Bayesian analysis uses an evidence based approach within the estimation of model parameters. This allows it to calculate certainty and uncertainty in the parameter estimates of custom built models conditional on the volume of evidence supporting the model’s conclusions. Furthermore, the use of prior distributions allows the model to understand likely parameter values before the model fitting process begins. In a low data availability sport, this prior understanding is extremely advantageous compared to machine learning approaches, which randomly initialise parameter values and thus require more data to provide accurate parameter estimates. A Mixture Model is a probabilistic model, which is comprised of a set of mixture components and weights. The weights represent the relationship between the data and the mixture components, and provide an alternative method of aggregating a rugby league pitch’s spatial data around a set of mixture components (or centres), rather than within a set of zones. Furthermore, these mixture components are able to estimate multiple categorical values concurrently providing a methodology through which individual possession outcome probabilities could be estimated in rugby league.

As a consequence of the limitations of the MRP approach currently employed to produce EPV models within low data availability sports, the primary aim of this study was to introduce a novel Bayesian Mixture Model approach to the development of an EPV model in rugby league, which could a) produce a smooth pitch surface, and b) calculate individual possession outcome probabilities. A secondary aim of the study was to show how the model could be used to identify differences in teams’ attacking and defending performances and evaluate player performances.

\section{Methodology}
Event level match-play data were obtained from Opta (Stats Perform, London, UK) for all 138 matches of the 2021 Super League season. In total, 557,050 match events were recorded, covering a range of actions (e.g. passes, kicks and runs) and descriptive data (e.g. video referee reviews, yellow and red cards). Across the season, 1001 tries were scored (768 successful conversion kicks, 233 unsuccessful conversion kicks), 175 penalty goals were attempted (158 successful, 17 unsuccessful) and 83 drop goals were attempted (37 successful, 46 unsuccessful).Prior to analysis, informed consent was obtained and ethics approval was provided by a university sub-ethics committee.

\subsection{Data Preprocessing} \label{ss:preprocessing}
For the purposes of this study, only actions performed by the attacking team were required. To facilitate the data filtering process, actions were split into 23 preprocessing categories (Table \ref{t:preprocessing_categories}). From these 23 categories, only actions belonging to the ``move team"; ``move self"; ``catch kick"; ``kick position"; ``move team error"; ``move self error"; ``loose ball" and ``kick goal" categories were used. Using only this subset of actions was necessary as a player completing a ``move self" action could also be given a ``run action" descriptor if they ran past an opponent, thus providing two entries for the same action (i.e. a multiple action coding). Additionally, incomplete/unsuccessful actions from any of these categories were only included if they resulted in the end of a possession. For example, an unsuccessful attempt at a loose ball collection or an unsuccessful attempt to intercept a pass by the defensive team were not included as these events were not successfully completed and including them would have unduly affected the chain of possession. However, an incomplete pass, or a dropped catch from a pass by the attacking team were included as these actions resulted in a change of possession.

\begin{table}
\tbl{Preprocessing categories and actions included within them in this study}
{\begin{tabular}{p{0.2\textwidth}p{0.7\textwidth}} \toprule
 Category & Events \\ \midrule
Auxiliary Information & Front Marker, Back Marker, Video Ref, Interchange, HIA, Stoppage\\
Generic Descriptor & Other Error, Try cause, Line Break Involvement, Line Break Assist, Break Cause, Tackle Break, Opp Error, Passing Move, Close Range, Error, Try Involvement, Long Range, Individual Effort, Other, Sin Bin Out, Yellow, Sin Bin Return To Field, Grounding, Contest, Sent Off Out, Red, Touchline/Deadball, Onside, On Report\\
Restart Actions & 50m Restart, Goal Line Drop Out, 20m Restart\\
Move Self & Restart Run, Evasion, Hitup, Kick Return, Line Engaged, Ruck Run, Dummy Half, Run, Line Not Engaged\\
Move Team & Complete, Short - Crossfield, Short - Grubber, Own Player, Break, Short - Banana, To Ground, Short - Bomb, Short - Chip, Try\\
Kick Goal & Conversion, Penalty Goal, Field Goal\\
Kick Position & Long - To Opposition, Good, Long - To Open, Long - 40-20, Long - Touch\\
Catch Pass & Simple Receipt, Jump Catch\\
Catch Kick & Kick Receipt, Restart Receipt\\
Loose Ball & Defensive Cleanup, Attacking Cleanup, Attempted Intercept, Contestable Cleanup, Interception\\
Tackle & Made, Dominant, Offload To Ground, Turnover Ball Split, Forced Within In Goal, Stolen, Turnover Into Touch, Offload\\
Missed Tackle & Bumped Off, Stepped, Positional, Outpaced, Try Conceded\\
Run Action & Dummy Pass, Half Break, Line Break, Carried Dead Ball, Forced Into In Goal, Carried In Touch\\
Play-The-Ball & Lost, Won, Interrupted \\
Attacking Descriptor & Kick Line Break, Try Assist, From Kick, From Penalty, From Line\\
Defensive Descriptor & Kick Pressure\\
Move Self Error & Dropped Ball Unforced, Ball Jolted, Lost Ball Forced \\
Move Team Error & Not Out, Failure To Find Touch, Incomplete, Off Target, Forward Pass, Forward, Kick Error, Bad Offload, To Opposition, Bad Pass, PTB Fumble, Intercepted\\
Catch Error & Accidental Knock On, Falcon\\
Penalty Conceded & Defence, Penalty, Inside 10m, Attack, Ruck Infringement, Foul Play, Double Movement, Obstruction\\
Defensive Play & Flop, Kick Not Defused, Kick Defused, Charge Down, Attempted Steal\\
Off The Ball & Decoy, Support Run, Kick Shield, Kick Shepherd, Kick Chase\\ \bottomrule
\end{tabular}}
\label{t:preprocessing_categories}
\end{table}

In this study, the location of the action was used for analysis. Consequently, it was important that locations were not coded on more than one occasion for different actions completed consecutively within the same sequence. For example, if a player caught the ball and decided to run in the same location, Opta would code each action as an individual observation. As the type of action was not important for this study, these multiple location codings were removed from the dataset indiscriminately by deleting the second consecutive location observation for a player in the same episode. The filtering of data and removal of multiple action/location codings described above resulted in a final dataset of 99,966 actions. 

To enable the estimation of individual possession outcome probabilities, possessions and possession outcomes were defined. A possession began when a team successfully gained possession of the ball and ended due to a handover, loss of possession caused by an error/foul play, points being scored or a goal kick attempt. It was therefore possible for an attacking possession to encompass more plays than the typical attacking set of 6 tackles if an error/foul was made by the opposition team. Five possession outcomes were defined and treated as discrete categories: converted try; unconverted try; penalty goal; drop goal; and no try. These possession outcomes are the same as those used in previous studies (\cite{Kempton2016,Sawczuk2021}). However, by treating them discretely it was possible to estimate individual probabilities, which differs from the previous approach of treating them continuously and aggregating the results into a single value (\cite{Kempton2016,Sawczuk2021}).

The data were organised into 25 subsets to allow the estimation of possession outcome probabilities at two levels: whole league; and team attacking/defensive. The whole league data was represented by all 99,966 observations. This model provided an understanding of the average possession outcome probabilities for each $x,y$ location across all teams in the league. The team attacking and defensive models provided the probability of a single team scoring (attacking models) or conceding (defensive models) possession outcomes for every $x,y$ location on the pitch. 12 subsets were created to model each team's attacking data, using only actions performed by a given team (median 8105 actions per team, interquartile range 7596-8937); 12 further subsets were created to model each team's defending data, using only actions performed by all teams against a specific team (median 8077 actions per team, interquartile range 7878-8700). Table \ref{t:sample_possession} provides a sample possession from the dataset.

\begin{table}
\tbl{Sample possession used in this study. Data includes the teams involved in the possession, the player ID, the $x,y$ coordinates of the action, the possession number (PosNum) and the possession outcome (PosCat; in this case no try for all rows).}
{\begin{tabular}{ccccccc} \toprule
Attacking Team & Defending Team & Player ID & $x$ & $y$ & PosNum & PosCat\\ \midrule
St Helens & Salford & 3107 & 9 & 4 & 1 & 0\\
St Helens & Salford & 21716 & 9 & 6 & 1 & 0\\
St Helens & Salford & 1983 & 14 & 11 & 1 & 0\\
St Helens & Salford & 2904 & 22 & 13 & 1 & 0\\
St Helens & Salford & 11439 & 12 & 12 & 1 & 0\\
St Helens & Salford & 21795 & 37 & 16 & 1 & 0\\
St Helens & Salford & 20567 & 36 & 24 & 1 & 0\\
St Helens & Salford & 2904 & 54 & 35 & 1 & 0\\ \bottomrule
\end{tabular}}
\label{t:sample_possession}
\end{table}

\subsection{A Bayesian Mixture Model Approach}
A Bayesian Mixture Model was used to develop the novel EPV model, which provided a smooth pitch surface and individual possession outcome probabilities in rugby league. The model estimates the probability of each possession outcome ($s \in S$) for a set of mixture components (or 'centres') on the pitch. Each centre holds probabilities for the five possession outcomes, represented as a 5-dimensional vector. The probabilities for any $x,y$ location on the pitch are calculated by taking a weighted average of probabilities at all centres on the pitch using a set of weights $Z$. Denoting by $P(s; x,y)$ the probability of possession outcome $s$ at location $(x,y)$ is calculated as
\begin{equation}
P(s; x,y) = \sum_k z_k(x,y) P_k(s)
\label{eq:psxy}
\end{equation}
where $z_k(x,y)$ is the weight corresponding to the location $(x,y)$ and $k$-th centre and $P_k(s)$ is the probability of possession outcome $s$ at the centre $k$. 

Unlike some Mixture Models, which estimate the weights for each centre as part of the model, the weights in this model are treated as fixed for every $x,y$ location on the pitch (Section \ref{ss:centre_weights}). With these fixed weights, Bayesian analysis is used to compute the distribution of possession outcome probabilities at all centres $P_k$. The prior distribution for $P_k$ is Dirichlet with the parameter, $\alpha$: 
\[
P_k \sim \text{Dirichlet} (\alpha).
\]
The prior distributions are independent between centres (i.e. there were separate prior distributions, denoting the prior probabilities of all five possession outcomes occurring, for centres 1, 2, 3, 4 etc.). The Dirichlet distribution is a multivariate generalisation of the beta distribution, parameterised by a vector of positive reals ($\alpha$). Its output is a vector of non-negative numbers summing up to one, i.e., a vector of probabilities. As it is only possible for one possession outcome to occur every possession in rugby league, this distribution was ideally suited as the prior for possession outcome probabilities.

The likelihood function for a set of data, $D = (x_i, y_i, s_i)_{i=1}^n$ is given by:
\[
P(D | \boldsymbol{\alpha}) = \prod_{i=1}^n  \sum_k z_k(x_i, y_i) P_k(s_i|\alpha_k),
\]
where $\boldsymbol{\alpha} = (\alpha_k)_k$ is the vector of centre Dirichlet parameters and
\[
P_k(s |\alpha_k) = \int \pi_s p_{\text{Dirichlet}} (p |\alpha_k) d\pi,
\]
where $p_{\text{Dirichlet}} (\pi |\alpha_k)$ is the density of Dirichlet distribution with parameter $\alpha$ at point $\pi \in \{(\pi_1, \ldots, \pi_5) \in [0,1]^5:\ \pi_1 + \ldots + \pi_5 = 1 \}$.

With the above notation, the prior is $\alpha_k = \alpha$ for all $k$; the posterior distribution is the distribution of $p(\boldsymbol(\alpha) | D)$. It is intractable as each observation influences 4 centres (in the field of play) or 2 centres (in the try area), and the weights and centres vary between observations. Therefore, Markov Chain Monte Carlo (MCMC) sampling was used to estimate the probabilities at each centre. MCMC sampling methods allow for the systematic random sampling from high dimensional probability distributions and those samples were used to draw conclusions on the posterior distribution of probabilities on the pitch. 

\subsection{Centre Weights} \label{ss:centre_weights}
\begin{figure}
\centering
\includegraphics[height=0.5\textheight]{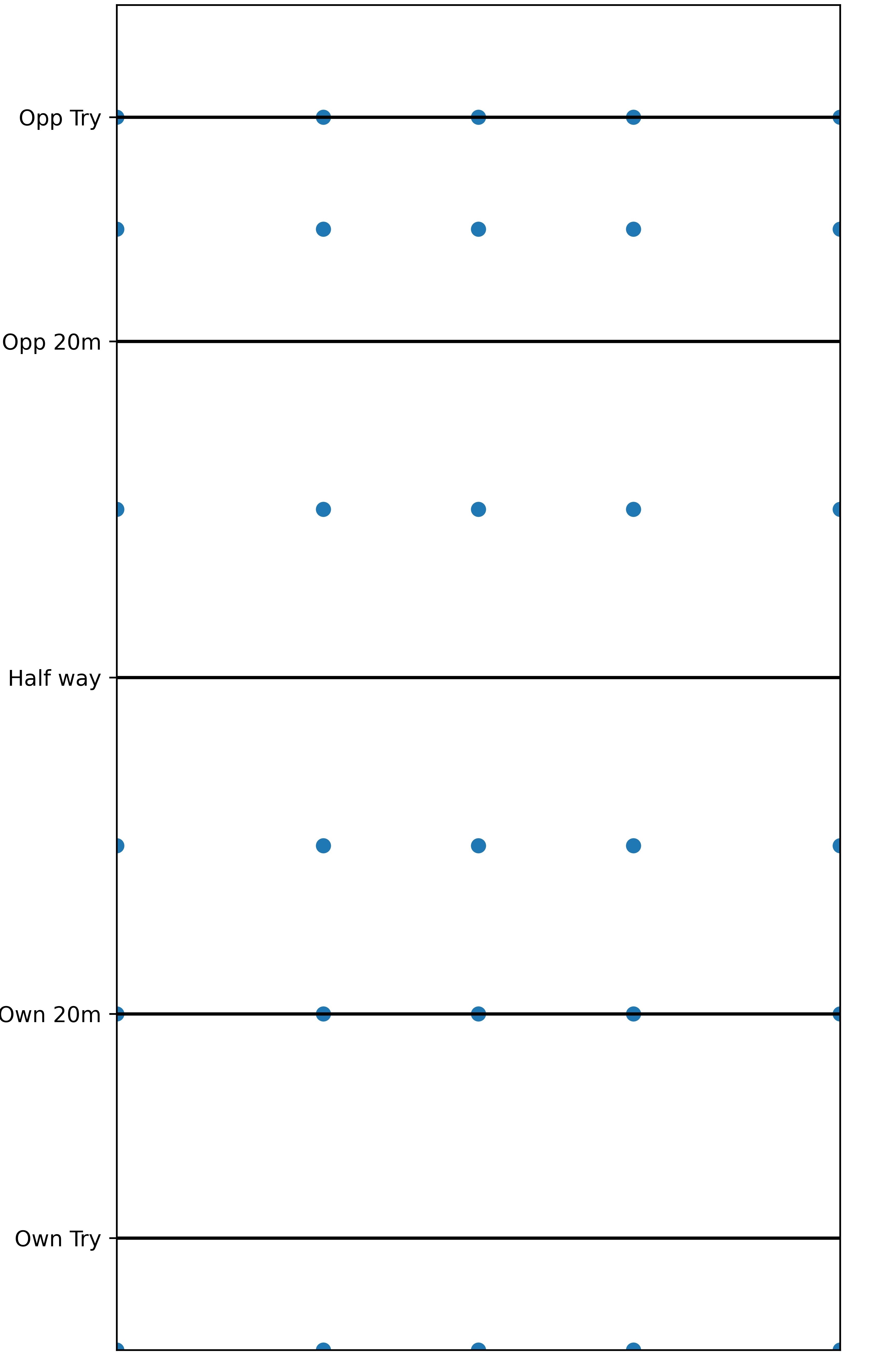}
\caption{Location of the 30 field of play centres used in this study. Three centres (not plotted) were included in the try area at $x$ $\in$ \{0, 35, 70\}, equivalent to the left, middle and right centres in the field of play. No $y$ coordinate was considered for try area centres.} \label{fig:centre_locations}
\end{figure}

After consultation with professional experts, 33 centres were placed around the pitch. 30 centres were located in the field of play, uniformly positioned at $x$ $\in$ \{0, 20, 35, 50, 70\} and $y$ $\in$ \{-10, 20, 35, 65, 90, 100\}. These locations were chosen to ensure that every location within the field of play would fall within a ``zone" defined by four centres. Figure \ref{fig:centre_locations} plots the centre locations for the field of play. 3 centres were located in the opposition team try area ($x$ $\in$ \{0, 35, 70\}). No $y$ coordinate was considered for centres in the try area as the actions players choose in this area are not influenced by their $y$ coordinate. The field of play and try area centres were evaluated separately. This decision was made due to the different player behaviours that are observed in the two areas: in the field of play, players are equally likely to choose different actions dependent on the game situation; in the try area, players will attempt to ground the ball for a try as soon as possible irrespective of the game situation. 

Each $x,y$ location on the pitch was assigned 33 weights, which described the aggregation of their data to the centres in the model. 30 weights were calculated in the field of play using bilinear interpolation;, 3 weights were calculated in the try area using linear interpolation. In line with the assumption of independence between the two areas of the pitch, any location in the field of play was automatically given weights of $0$ for the try area centres; any location in the try area was automatically given weights of $0$ for the field of play centres.

Locations in the field of play had a maximum of four non-zero weights. The value of the weights for each $x,y$ location in the field of play was derived from the distance between the $x,y$ location and the four centres surrounding it in a quadrilateral shape. These centres had coordinates ($x_1$, $y_1$), ($x_1$, $y_2$), ($x_2$, $y_1$) and ($x_2$, $y_2$). The weights ($z_{11}$, $z_{12}$, $z_{21}$ and $z_{22}$) of these centres for location $x,y$ were calculated using Equations \ref{eq:z11_blint_weights}, \ref{eq:z12_blint_weights}, \ref{eq:z21_blint_weights} and \ref{eq:z22_blint_weights}. The remaining centres were assigned a weight of $0$. 
\begin{equation}
    z_{11} = \frac{(x_2 - x)(y_2 - y)}{(x_2-x_1)(y_2-y_1)},
    \label{eq:z11_blint_weights}
\end{equation}
\begin{equation}
    z_{12} = \frac{(x_2 - x)(y - y_1)}{(x_2-x_1)(y_2-y_1)},
    \label{eq:z12_blint_weights}
\end{equation}
\begin{equation}
    z_{21} = \frac{(x - x_1)(y_2 - y)}{(x_2-x_1)(y_2-y_1)},
    \label{eq:z21_blint_weights}
\end{equation}
\begin{equation}
    z_{22} = \frac{(x - x_1)(y - y_1)}{(x_2-x_1)(y_2-y_1)}.
    \label{eq:z22_blint_weights}
\end{equation}

Locations in the try area had a maximum of two non-zero weights. Here, only the $x$ location of the centres ($x \in \{0, 35, 70\}$) was considered so the weights were derived from the distance between the $x$ coordinate of the action location and the $x$ coordinate of the centre. For an $x,y$ location in the try area, linear interpolation between the two closest centres $x_1$, $x_2$, with $x_2 > x_1$, from the above set of three, was used to calculate two weights $z_1$, $z_2$; the weight of the remaining centre was set to $0$. The non-zero weights were given by:
\begin{equation}
    z_{1} = \frac{x_2 - x}{x_2 - x_1},
    \label{eq:z0_lint_weights}
\end{equation}
\begin{equation}
    z_{2} = \frac{x - x_0}{x_2 - x_1}.
    \label{eq:z1_lint_weights}
\end{equation}

\subsection{EPV Calculation}
The EPV for a location was derived from the possession outcome probabilities. The EPV for a location with coordinates $x,y$ was calculated using the probability of the five possession outcomes and their true points scoring values:
\begin{equation}
    \text{EPV}_{(x,y)} = \sum_{s\in{S}}P(s; x,y)\text{Points}(s)
    \label{eq:epv_calculation}
\end{equation}
where $P(s; x,y)$ is the probability of possession outcome $s$ in location $x,y$ (Equation \ref{eq:psxy}) and Points($s$) is the true point scoring value of possession outcome $s$ (converted try = 6; unconverted try = 4; penalty goal = 2; drop goal = 1; no score=0).

\subsection{Modelling Procedure}
The analysis for this study was conducted at two levels. First, the model was run using the whole league data, then the 24 team attacking and defending models were run, using the data subsets described in Section \ref{ss:preprocessing}. It has previously been argued that a hierarchical approach is the most appropriate method of analysing data at multiple levels (\cite{vandeSchoot2021}), however adopting such a method would have only allowed a single averaged shift (either positive or negative) away from the whole league model estimates. As such, if a team had a greater probability of scoring a try from the left side of the pitch in the opposition try area, but a greater probability of scoring from the right side of the pitch in their own half than the whole league model estimates, it would not be possible to identify these differences using a hierarchical approach. In this study, completing two levels of analyses (i.e. whole league, then team level) allowed the team level models to include information already learned about the league within their parameter estimates through the use of an $\alpha$ prior distribution estimated from the whole league model posterior distribution.

The whole league data model provided league average possession outcome probabilities. For this proof of concept model, human-defined priors were used for $P_k$ (Appendix \ref{app:whole_league_priors}). These priors were selected after discussion with experts and were informed by previous research (\cite{Kempton2016, Sawczuk2021}). They loosely informed the model that there was a greater chance of points being scored by the end of the possession the closer the location was to the opposition try line. 

24 models were produced at the team level, one for each team's attacking and defending data subset. Maximum likelihood estimation of the posterior distribution of the whole league model was used to calculate the $\alpha$ priors for these analyses (Appendix \ref{app:team_priors}). All team models used the same $\alpha$ prior distribution.

The mean of the posterior distribution of possession outcome probabilities $P_k^\mu$ for each model was substituted into Equations \ref{eq:psxy} and \ref{eq:epv_calculation} to provide individual possession outcome probabilities and EPV for each $x,y$ location:
\[
P^\mu(s; x,y) = \sum_k z_k(x,y) P_k^\mu(s)
\]
\[
\text{EPV}_{(x,y)}^\mu = \sum_{s\in{S}}P^\mu(s; x,y)\text{Points}(s)
\]

The standard deviation of the posterior distribution of possession outcome probabilities $P_k^\sigma$ for each model was used to provide an understanding of uncertainty within the parameter estimates for any $x,y$ location. This required Equations \ref{eq:psxy} and \ref{eq:epv_calculation} to be modified:
\[
P^\sigma(s; x,y) = \sqrt{\sum_k z_k(x,y) P_k^\sigma(s)^2}
\]
\[
\text{EPV}_{(x,y)}^\sigma = \sqrt{\sum_{s\in{S}}P^\sigma(s; x,y)^2\text{Points}(s)}
\]

A smooth pitch surface was produced by calculating these values for every $x,y$ location on the pitch. Smooth pitch surfaces were produced for each of the five possession outcomes and EPV for the whole league model and every team attacking and defending model.

\subsection{Analysing Team and Player Performances}
To showcase potential uses for the model, methods of understanding team and player performances are provided. Team performances were evaluated using visual inspection of the smooth pitch plots for each possession outcome and the EPV model. The plots compare the probabilities/EPV of each team attacking/defending model to the whole league model and provide an understanding of areas where a team is more or less likely to generate value at the individual possession outcome or overall EPV level on the pitch compared to the average team.

Player performances were evaluated using Actual vs Expected (AE) player performance ratings, devised from the EPV values generated by the whole league model. Player ratings were calculated via Equation \ref{eq:player_ratings} and compared the actual points return of a possession to the expected points return using $\text{EPV}_{(x,y)}^\mu$. The sum of the differences between actual and expected returns was divided by the median number of possessions per player team per fixture. This adjustment ensured that players from teams who had more possessions within a match were not unduly favoured by the results.

\begin{equation}
    \text{Player AE rating} = \frac{\text{Actual return} - \text{Expected Return}}{\text{Player team median number of possessions per fixture}},
    \label{eq:player_ratings}
\end{equation}

All preprocessing and analysis was completed using bespoke Python scripts (Python 3.7, Python Software Foundation, Delawere, USA) and the PyMC3 v3.11.4 package \cite{pymc}.

\section{Results}
\subsection{Whole League Model}
Of the 99,966 actions included in this study, only 91 occurred in the try area. The $\text{EPV}_{(x,y)}^\mu$ of these actions was much greater than those actions outside of the try area. For example, using centre values, the highest EPV in the field of play was at centre $(50,100)$, where $\text{EPV}_{(50,100)}^\mu = 1.73$. All three try area centres had much greater values ($\text{EPV}_{(x,y)}^\mu \in \{3.52, 3.72, 3.16\}$). For clarity of figure presentation, the try area values are removed from all plots. 

\begin{figure}
\centering
\includegraphics[width=0.8\textwidth]{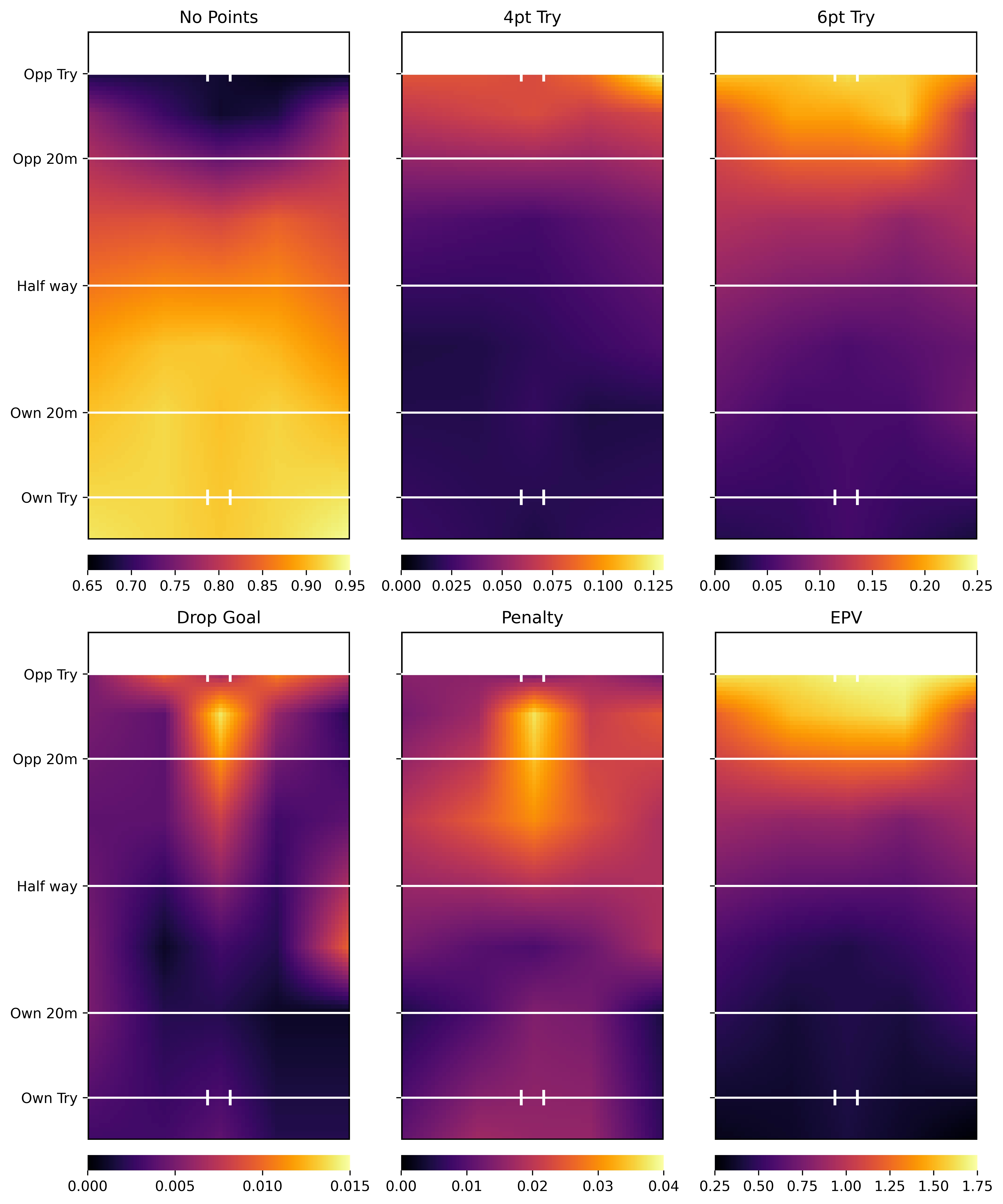}
\caption{Whole league model mean plot. 4pt Try and 6pt Try refer to unconverted and converted tries respectively. With the exception of EPV, probabilities are plotted. Smooth pitch surface for possession outcome probabilities is calculated using Equation \ref{eq:psxy} for each $x,y$ location on the pitch. EPV for each location is calculated using Equation \ref{eq:epv_calculation}. Brighter areas represent higher values.} \label{fig:wl_epv_mean}
\end{figure}

Figure \ref{fig:wl_epv_mean} shows the smooth pitch surface provided by the whole league model mean parameter estimates. There is a greater probability of points being scored by the end of the possession, the closer the location is to the opposition try area. The darker arc on the no points probability surface close to the opposition try line indicates that more points are likely to be scored from central locations than wider locations unless a player is extremely close to the try line. In general, a converted try was more likely to occur at the end of a possession than an unconverted try and a penalty goal was more likely to be scored than a drop goal.

\begin{figure}
\centering
\includegraphics[width=0.8\textwidth]{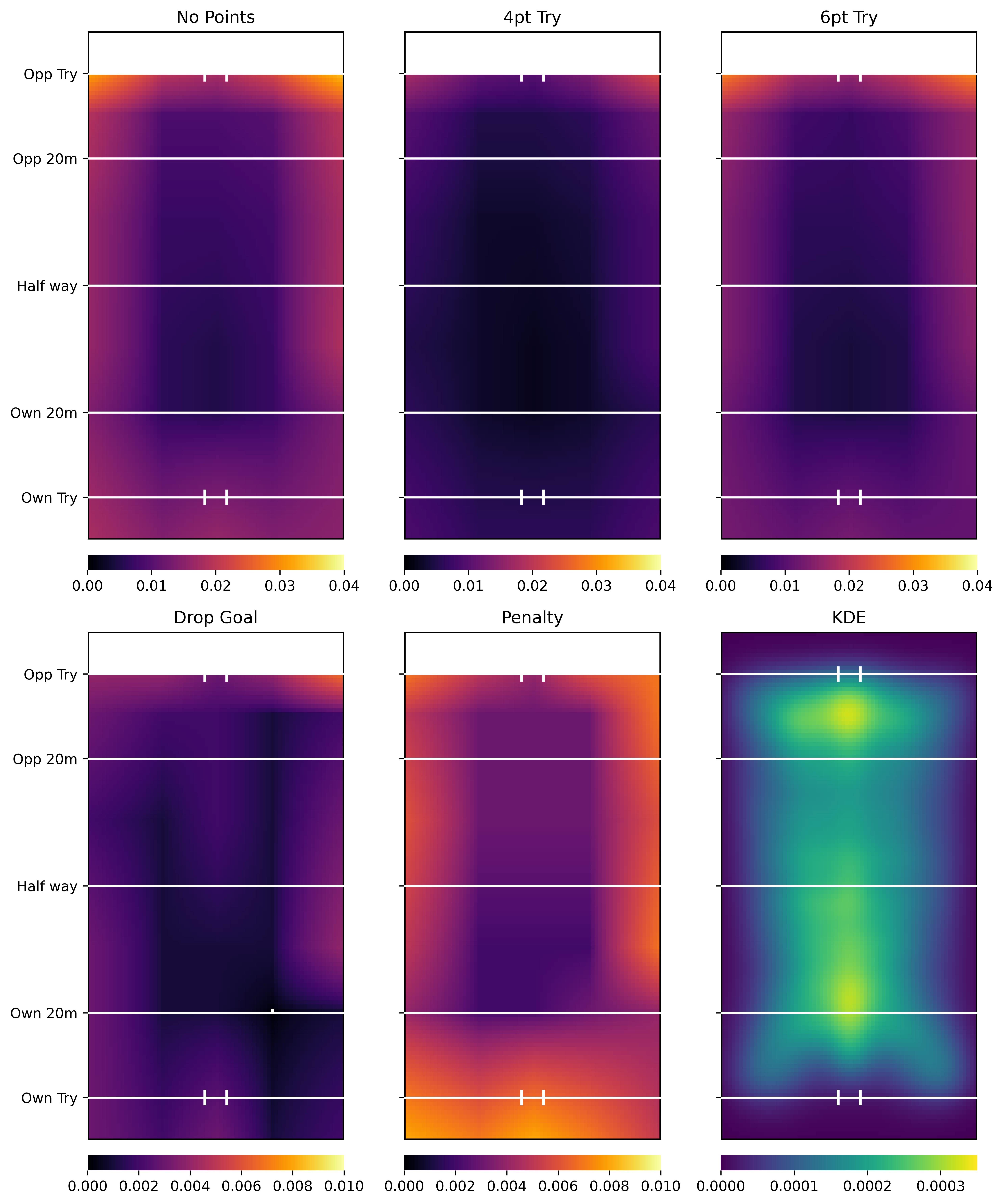}
\caption{Whole league model EPV standard deviation plot. 4pt Try and 6pt Try refer to unconverted and converted tries respectively. Kernel Density Estimation plot of dataset is provided to give an understanding of location densities. Brighter colours indicate higher values.} \label{fig:wl_epv_sd}
\end{figure}

Figure \ref{fig:wl_epv_sd} provides the variability in the smooth pitch surface using the whole league model parameter estimate standard deviations. In all possession outcome plots, there is greater variability in the wider areas of the pitch, which is accompanied by a lower density of actions in those areas in the KDE plot. This variability is particularly large for the penalty goal probabilities in wide areas. Similarly, there is increased variability in try/no try probabilities in both corners of the pitch on the opposition try line. 

\subsection{Team Attacking and Defending Models}
\begin{figure}
\centering
\includegraphics[width=0.8\textwidth]{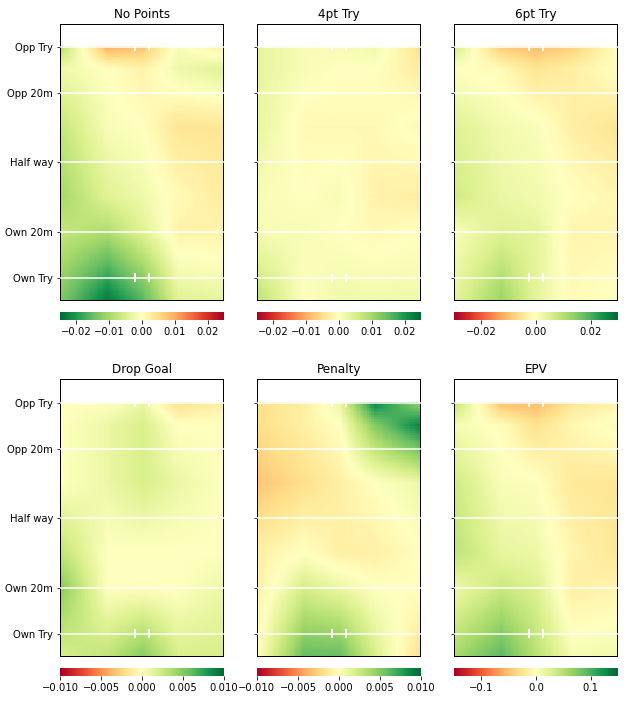}
\caption{Team A smooth pitch surface plot from team attacking model. Green areas represent higher value for a more favourable possession outcome (i.e. greater probability of all events occurring except No Points) compared to whole league model. 4pt Try and 6pt Try refer to unconverted and converted tries respectively.} \label{fig:teama_att}
\end{figure}

\begin{figure}
\centering
\includegraphics[width=0.8\textwidth]{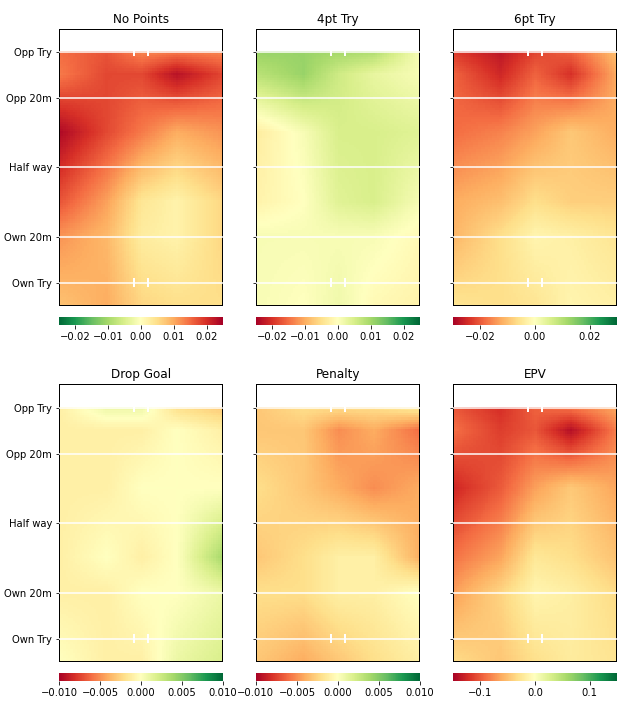}
\caption{Team B smooth pitch surface plot from team attacking model. Green areas represent higher value for a more favourable possession outcome (i.e. greater probability of all events occurring except No Points) compared to whole league model. 4pt Try and 6pt Try refer to unconverted and converted tries respectively.} \label{fig:teamb_att}
\end{figure}

Figures \ref{fig:teama_att} (Team A) and \ref{fig:teamb_att} (Team B) provide the attacking pitch surface plots from two teams' attacking models. There are clear differences between the two plots in different areas across the pitch for all possession outcomes. For example, Team A have much higher value on the left side of the pitch in all plots except penalty goals. This is mainly shown by a reduced probability of no try, increased probability of a converted (6pt) try and drop goals on the left side of the pitch. The EPV plot shows the trend even more clearly. Conversely, Team B showed below average attacking prowess. The red areas in all plots except the unconverted (4pt) try indicate that they were more likely to score no points by the end of their possessions and less likely to score any type of points outcome than the average team except unconverted tries.

\begin{figure}
\centering
\includegraphics[width=0.8\textwidth]{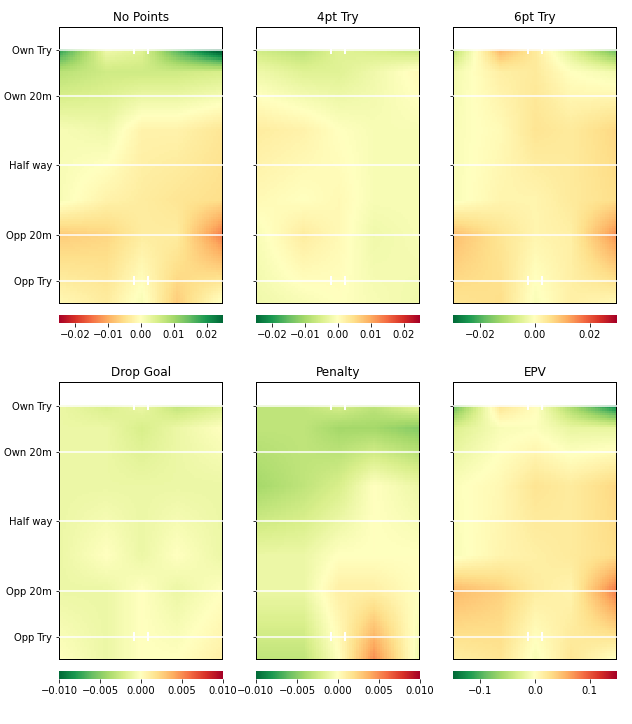}
\caption{Team A smooth pitch surface plot from team defending model. Green areas represent higher value for more favourable event (i.e. greater probability of all events occurring except No Points) compared to whole league model. 4pt Try and 6pt Try refer to unconverted and converted tries respectively.} \label{fig:teama_def}
\end{figure}

\begin{figure}
\centering
\includegraphics[width=0.8\textwidth]{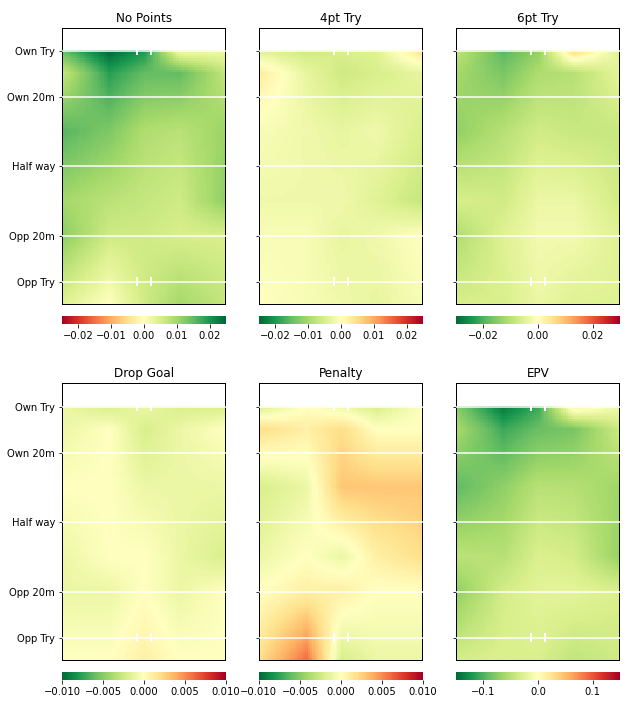}
\caption{Team B smooth pitch surface plot from team defending model. Green areas represent higher value for more favourable event (i.e. greater probability of all events occurring except No Points) compared to whole league model. 4pt Try and 6pt Try refer to unconverted and converted tries respectively.} \label{fig:teamb_def}
\end{figure}

Figures \ref{fig:teama_def} (Team A) and \ref{fig:teamb_def} (Team B) provide the defending plots from the same two teams' defending models. Clear differences between the two teams are again visible. Team A are have a greater probability than the average team of conceding points up until their own 20m line (shown by the reddish areas on the EPV plot and the no points plot), but have a reduced probability of conceding any points outcome on their own try line in the left and right corners of the pitch. Team B are considerably better than average at defending in all areas except the probability of the opposition team scoring penalty goals.

\subsection{Player Ratings}
Table \ref{t:player_ae} provides player ratings for the 20 best players in the 2021 Super League season according to the AE ratings. A selection of traditional summary statistics (tries, try assists, metres and goals kicked) are provided so the player AE ratings can be compared to these readily available statistics. The Man of Steel and the Young Player of the Year were both present within the top-20 players. Furthermore, a wide range of positions were present within the top-20 players, as well as a wide range of scoring profiles based on the four traditional summary statistics provided.

\begin{table}
\tbl{Top 20 player ratings as assessed by the AE ratings (Equation \ref{eq:player_ratings}). Tries, Try Assists, Metres and Goals are provided as references of statistics currently provided for player performances. To protect anonymity, reference statistics are provided as: T-5 (accumulated count within the top 5 players); T-10 (within the top 10 players), T-20 (within top 20 players) and 20+ (outside top 20 players).}
{\begin{tabular}{ccccccc} \toprule
Player & Position & AE Rating & Tries & Try Assists & Metres & Goals\\ \midrule
276 & Full Back & 8.21 & T-20 & T-5 & 20+ & 20+ \\
19 & Winger & 6.67 & T-5 & 20+ & T-5 & 20+ \\
6335 & Stand-off & 6.35 & 20+ & 20+ & 20+ & T-5\\
1004 & Scrum Half & 6.10 & 20+ & 20+ & 20+ & 20+\\
433 & Full Back & 4.96 & 20+ & T-10 & 20+ & T-5\\
158 & Winger & 4.82 & T-5 & 20+ & T-10 & 20+ \\
188 & Centre & 4.78 & T-5 & 20+ & T-5 & 20+\\
1249 & Hooker & 4.37 & 20+ & 20+ & 20+ & 20+\\
92 & Scrum Half & 3.91 & 20+ & T-5 & 20+ & 20+\\
1281 & Loose Forward & 3.86 & 20+ & 20+ & 20+ & 20+\\
406 & Loose Forward & 3.82 & 20+ & 20+ & 20+ & 20+ \\
371 & Winger & 3.37 & T-20 & 20+ & T-20 & 20+ \\
8 & Prop & 3.33 & 20+ & 20+ & T-10 & 20+ \\
282 & Second Row & 3.28 & 20+ & 20+ & 20+ & 20+ \\
20528 & Winger & 3.11 & T-5 & 20+ & T-5 & 20+ \\
22852 & Full Back & 3.10 & T-10 & T-20 & 20+ & 20+ \\
5 & Winger & 3.08 & T-20 & 20+ & 20+ & 20+ \\
26 & Hooker & 2.96 & 20+ & 20+ & 20+ & 20+ \\
440 & Loose Forward & 2.88 & 20+ & 20+ & 20+ & 20+ \\
988 & Stand-off & 2.81 & 20+ & 20+ & 20+ & 20+ \\ \bottomrule
\end{tabular}}
\label{t:player_ae}
\end{table}

\section{Discussion}
The primary aim of this study was to introduce a novel Bayesian Mixture Model approach to the development of an EPV model in rugby league, which could a) produce a smooth pitch surface, and b) calculate individual possession outcome probabilities. A secondary aim of the study was to show how the model could be used to identify differences in teams’ attacking and defending performances and evaluate player performances.

\subsection{The Bayesian Mixture Model Approach}
The key contribution of this study is the adoption of a Bayesian Mixture Model approach to develop an EPV model which produced a smooth pitch surface of individual possession outcome probabilities. The Bayesian Mixture Model worked by assigning a set of centres across the pitch and using the proximity of a location to these centres to aggregate data, improving upon the zonal approaches previously employed (\cite{Kempton2016, Cervone2016, Sawczuk2021}). With the low data availability present in rugby league, such an approach was best modelled using an evidence-based Bayesian approach. The Bayesian approach allowed prior distributions to provide the model an understanding of possible parameter estimates before the fitting process began and allowed information sharing between the different levels of analysis (i.e. using the whole league model posterior distribution to calculate the team attacking/defending models' prior distributions).

Figures \ref{fig:wl_epv_mean} (mean) and \ref{fig:wl_epv_sd} (standard deviation) depict the smooth pitch surfaces provided by the whole league model. At the EPV level, the results are similar to previous studies, which suggest that there is greater value in areas closer to the opposition try line (\cite{Kempton2016, Sawczuk2021}) and more centrally (\cite{Sawczuk2021}). However, at the individual possession outcome level, much greater insights are obtained. For example, the importance of central areas to drop goal and penalty goal success is clearly shown. Similarly, there is an increase in the probability of unconverted tries being scored on the far right of the pitch. Neither of these insights would be possible without the estimation of individual possession outcome probabilities. Similarly, the standard deviation plot shows that there is a lot of certainty in parameter estimates in the centre of the pitch, but much less around the outer part of the pitch. This information allows practitioners to temper their confidence in the probabilities provided in these areas. This understanding of the variability of parameter estimates is not typically available when using machine learning methods showing the strength of the Bayesian model employed.

The Bayesian Mixture Model approach provides an exceptional amount of flexibility, both with respect to the EPV measure and the probabilities it could estimate. In this study, the EPV was defined as the weighted average of all possession outcomes, but future studies may wish to consider whether using try/no try probabilities can provide additional insights. Similarly, the mixture model approach can handle more probabilities if required. For example, by preprocessing the data in a different way and providing categories for the next scoring outcome (converted try, unconverted try, penalty goal and drop goal for both teams, alongside no score), it would be possible to estimate 9 scoring outcome probabilities and use these to develop a "next scoring outcome" EPV measure.

\subsection{Team Level Insights}
Figures \ref{fig:teama_att} and \ref{fig:teamb_att} provide attacking pitch surfaces for two separate teams. It is clear from visual inspection of the plots that different insights can be generated. For example, Team B were particularly poor at attacking across the pitch. Conversely, Team A was more likely to score tries on the left side of the pitch, but more likely to score a penalty goal on the right side of the pitch. Figures \ref{fig:teama_def} and \ref{fig:teamb_def} provide defensive pitch surfaces for the same two teams. Again, the plots provide different insights. It can be deduced from Figure \ref{fig:teama_def} that Team A are much more likely to concede points from possessions beginning outside their 20m. However, if actions occur close to their try line, they are able to defend them particularly well in the corners of the pitch. Team B were excellent defensively across the pitch. Indeed, the penalty goal plot suggests that opposition teams were more likely to try and score penalties against them than the average team, potentially because they were unable to break down the defence and score tries. The ability of the model to identify these differences between teams is extremely valuable with respect to developing tactical strategies for upcoming matches. It would now be beneficial to develop a methodology through which differences in these probabilities or values can be objectively identified.

\subsection{AE Player Ratings}
Table \ref{t:player_ae} provides the top 20 players across the 2021 Super League season based on the AE player ratings. The 
rating denotes the points contribution actions taken by a player provided per match (e.g. player 276's actions contributed 8.21 points per match to their team's overall points count). A wide variety of positions and traditional statistic profiles are included. Some of these players scored more tries, others were better at providing try assists or kicking goals; others did not excel in any of the summary statistics provided. This ability to understand valuable players from different positions and across different summary statistic profiles differs from previous research (\cite{Sawczuk2021-UKCI}), which predominantly valued players who attempted try scoring action. The top 20 players included the Man of Steel and Young Player of the Year providing it with strong face validity. However, those players who were involved in a large number of actions in possessions which resulted in no try being scored due to the location they began in (e.g. most of the scrum halves) were valued poorly. It may therefore be appropriate to consider different inputs to the centres, or the player ratings, which are able to value these players appropriately.

\subsection{Limitations and Future Directions}
The model described in this paper significantly advances the approaches used in previous studies in rugby league (\cite{Kempton2016, Sawczuk2021}). It provides a flexible methodology, which could used to generate EPV models in any sport where data is not readily available but is subject to two key limitations. The first of these is that it only considers event level data, so there is limited context surrounding the value of the locations. Therefore, if a player is stood with 5 defenders directly in front of him or no defenders directly in front of him, his location would be valued the same. Secondly, the model does not yet incorporate the auto-correlation present within possession sequences. Although the impact of this on parameter estimates is likely to be limited due to the length of the possession sequences, it is still a limitation of the model. Alongside the future directions indicated in the sections above, future studies may wish to address these limitations when appropriate data is available.

\section{Conclusion}
In this paper, a novel Bayesian Mixture Model approach to the estimation of an EPV Model was proposed in rugby league. The model was able to provide a smooth pitch surface and estimate the probability of individual possession outcomes occurring. Insights into player and team performances were derived from the model showcasing its ability to provide valuable information for upcoming fixtures and player recruitment. Given appropriate data preprocessing and modification of possession outcome categories, the model could be adapted to any invasion sport's requirements.

\section*{Disclosure Statement}
No potential conflict of interest was reported by the authors

\bibliographystyle{abbrvnat}
\bibliography{BayesianJSS}

\pagebreak
\appendix
\section{Whole league model priors} \label{app:whole_league_priors}
Table of $\alpha$ values representing the prior distribution for each centre within the whole league model.
\begin{table}[h]
\tbl{Prior $\alpha$ values for whole league Bayesian model. Centre coordinates are provided, alongside $\alpha$ values for each possession outcome. C.Try refers to converted try (6 points) and U.Try refers to unconverted try (4 points).}
{\begin{tabular}{cccccc} \toprule
Centre & No Try & Drop Goal & Penalty & U.Try & C.Try\\ \midrule
( 0, -10) & 90 & 1 & 1 & 4 & 4 \\
(20, -10) & 90 & 1 & 1 & 4 & 4 \\
(35, -10) & 90 & 1 & 1 & 4 & 4 \\
(50, -10) & 90 & 1 & 1 & 4 & 4 \\
(70, -10) & 90 & 1 & 1 & 4 & 4 \\
( 0,  20) & 90 & 1 & 1 & 4 & 4 \\
(20,  20) & 90 & 1 & 1 & 4 & 4 \\
(35,  20) & 90 & 1 & 1 & 4 & 4 \\
(50,  20) & 90 & 1 & 1 & 4 & 4 \\
(70,  20) & 90 & 1 & 1 & 4 & 4 \\
( 0,  35) & 85 & 1 & 3 & 5 & 6 \\
(20,  35) & 85 & 1 & 3 & 5 & 6 \\
(35,  35) & 85 & 1 & 3 & 5 & 6 \\
(50,  35) & 85 & 1 & 3 & 5 & 6 \\
(70,  35) & 85 & 1 & 3 & 5 & 6 \\
( 0,  65) & 80 & 1 & 3 & 7 & 9 \\
(20,  65) & 80 & 1 & 3 & 7 & 9 \\
(35,  65) & 80 & 1 & 3 & 7 & 9 \\
(50,  65) & 80 & 1 & 3 & 7 & 9 \\
(70,  65) & 80 & 1 & 3 & 7 & 9 \\
( 0,  90) & 75 & 1 & 3 & 9 & 12 \\
(20,  90) & 75 & 1 & 3 & 9 & 12 \\
(35,  90) & 75 & 1 & 3 & 9 & 12 \\
(50,  90) & 75 & 1 & 3 & 9 & 12 \\
(70,  90) & 75 & 1 & 3 & 9 & 12 \\
( 0, 100) & 70 & 1 & 3 & 10 & 15 \\
(20, 100) & 70 & 1 & 3 & 10 & 15 \\
(35, 100) & 70 & 1 & 3 & 10 & 15 \\
(50, 100) & 70 & 1 & 3 & 10 & 15 \\
(70, 100) & 70 & 1 & 3 & 10 & 15 \\
( 0, TRY) & 35 & 1 & 1 & 28 & 35 \\
(35, TRY) & 35 & 1 & 1 & 28 & 35 \\
(70, TRY) & 35 & 1 & 1 & 28 & 35 \\ \bottomrule
\end{tabular}}
\label{t:whole_league_priors}
\end{table}

\pagebreak
\section{Team model priors} \label{app:team_priors}
Table of $\alpha$ values representing the prior distribution for each centre within the team attacking and defending models. All 24 models used the same prior distribution.

\begin{table}[h]
\tbl{$\alpha$ prior distribution values for team attacking and defending models. $\alpha$ values calculated via maximum likelihood estimation from posterior distribution of the whole league model. Centre coordinates are provided, alongside $\alpha$ values for each possession outcome. C.Try refers to converted try (6 points) and U.Try refers to unconverted try (4 points). Values rounded to 2 decimal places for brevity.}
{\begin{tabular}{cccccc} \toprule
Centre & No Try & Drop Goal & Penalty & U.Try & C.Try\\ \midrule
( 0, -10) & 184.72 & 0.72 & 2.01 & 4.97 & 6.43 \\
(20, -10) & 345.31 & 1.07 & 6.38 & 7.45 & 14.88 \\
(35, -10) & 280.81 & 1.37 & 4.63 & 5.08 & 16.00 \\
(50, -10) & 365.08 & 0.95 & 6.30 & 7.90 & 15.67 \\
(70, -10) & 230.12 & 0.77 & 1.28 & 5.40 & 6.11 \\
( 0,  20) & 371.56 & 2.28 & 1.87 & 7.08 & 27.40 \\
(20,  20) & 1559.21 & 2.73 & 14.96 & 28.85 & 86.65 \\
(35,  20) & 2815.84 & 6.64 & 44.81 & 67.00 & 168.66 \\
(50,  20) & 1599.20 & 1.03 & 22.07 & 25.46 & 94.59 \\
(70,  20) & 393.10 & 0.73 & 1.75 & 6.92 & 35.35 \\
( 0,  35) & 373.60 & 2.10 & 5.30 & 6.51 & 33.39 \\
(20,  35) & 1707.17 & 2.05 & 18.37 & 30.07 & 121.44 \\
(35,  35) & 2636.37 & 8.50 & 24.50 & 59.31 & 162.86 \\
(50,  35) & 1673.02 & 3.58 & 21.79 & 44.56 & 116.69 \\
(70,  35) & 310.63 & 3.68 & 6.61 & 10.50 & 25.85 \\
( 0,  65) & 439.66 & 2.03 & 10.38 & 17.21 & 63.22 \\
(20,  65) & 2189.67 & 10.97 & 65.76 & 75.37 & 294.36 \\
(35,  65) & 2546.81 & 26.14 & 94.46 & 84.30 & 355.25 \\
(50,  65) & 1726.64 & 6.24 & 49.69 & 68.55 & 197.80 \\
(70,  65) & 360.60 & 1.82 & 8.02 & 18.25 & 51.31 \\
( 0,  90) & 413.37 & 2.70 & 6.90 & 35.62 & 83.31 \\
(20,  90) & 1356.56 & 8.21 & 32.70 & 139.25 & 384.54 \\
(35,  90) & 2150.80 & 43.80 & 116.69 & 241.11 & 636.41 \\
(50,  90) & 1437.66 & 12.84 & 43.18 & 142.16 & 463.60 \\
(70,  90) & 365.55 & 0.91 & 11.60 & 34.77 & 50.19 \\
( 0, 100) & 154.63 & 1.15 & 3.38 & 18.26 & 50.01 \\
(20, 100) & 370.25 & 5.46 & 8.15 & 42.98 & 116.60 \\
(35, 100) & 427.75 & 3.94 & 7.62 & 47.37 & 146.33 \\
(50, 100) & 299.46 & 4.83 & 7.87 & 40.45 & 99.93 \\
(70, 100) & 121.38 & 1.30 & 2.36 & 24.68 & 34.07 \\
( 0, TRY) & 36.62 & 0.98 & 0.95 & 33.38 & 46.66 \\
(35, TRY) & 43.45 & 0.97 & 0.99 & 29.80 & 68.50 \\
(70, TRY) & 42.17 & 1.00 & 1.02 & 30.12 & 39.13 \\ \bottomrule
\end{tabular}}
\label{t:team_priors}
\end{table} 

\end{document}